\documentclass[10pt,journal,twoside,final]{IEEEtran}

\usepackage[noadjust]{cite}
\usepackage{graphicx,epstopdf}
\usepackage[table]{xcolor}
\usepackage{amsmath,amsfonts,amssymb,amsthm,mathtools}
\usepackage{array}
\usepackage{longtable,tabularx}
\usepackage{multirow}
\usepackage{booktabs}
\usepackage{algorithm}
\usepackage{algorithmic}

\usepackage{threeparttable}

\newtheorem{theorem}{Theorem}

\newtheorem{corollary}{Corollary}
\newtheorem{lemma}{Lemma}
\newtheorem{definition}{Definition}
\theoremstyle{remark}
\newtheorem*{remark}{Remark}

\DeclareMathOperator*{\sign}{sign}

\newcommand{\R}[1]{\mathbb{R}^{#1}}

\newcommand{\VEC}[3]{\begin{bmatrix} #1 \\ #2 \\ #3 \end{bmatrix}}
\newcommand{\HVEC}[3]{\begin{bmatrix} #1 & #2 & #3 \end{bmatrix}}

\newcommand{\A}{\mathcal{A}}
\newcommand{\B}{\mathcal{B}}
\newcommand{\C}{\mathcal{C}}

\newcommand{\s}{\sigma}

\newcommand{\TR}[2]{\mathcal{T}_{#1}\left(#2\right)}

\newcommand{\LBKP}{\boxtimes}

\usepackage{hyperref}
\hypersetup{
	colorlinks,
	linkcolor={red!80!black},
	citecolor={purple!80!black},
	urlcolor={blue!80!black}
}

\usepackage[caption=false,font=footnotesize]{subfig}
\usepackage{url}

\begin{document}
%
\title{Nonlinear Control Allocation Using A Piecewise Multi-Linear Representation}
%
%

\author{Jahanzeb~Rajput,~and~Hafiz~Zeeshan~Iqbal~Khan%
\thanks{J. Rajput, Centers of Excellence in Science and Applied Technologies, Islamabad 44000, Pakistan (e-mail: jahanzeb\_rajput@hotmail.com).}
\thanks{H. Z. I. Khan, Centers of Excellence in Science and Applied Technologies, Islamabad 44000, Pakistan (e-mail: zeeshaniqbalkhan@hotmail.com).}
}

\ifCLASSOPTIONpeerreview

\else
\fi



\maketitle

\begin{abstract}
Nonlinear control allocation is an important part of modern nonlinear dynamic inversion based flight control systems which require highly accurate model of aircraft aerodynamics. Generally, an accurately implemented onboard model determines how well the system nonlinearities can be canceled. Thus, more accurate model results in better cancellation, leading to the higher performance of the controller. In this paper, a new control system is presented that combines nonlinear dynamic inversion with a piecewise multi-linear representation based control allocation. The piecewise multi-linear representation is developed through a new generalization of Kronecker product for block matrices, combined with the canonical piecewise linear representation of nonlinear functions. Analytical expressions for the Jacobian of the piecewise multi-linear model are also presented. Proposed formulation gives an exact representation of piecewise multi-linear aerodynamic data and thus is capable of accurately modeling nonlinear aerodynamics over the entire flight envelope of an aircraft. Resulting nonlinear controller is applied to control of a tailless flying wing aircraft with ten independently operating control surfaces. The simulation results for two innovative control surface configurations indicate that perfect control allocation performance can be achieved, leading to better tracking performance compared with ordinary polynomial-based control allocation.
\end{abstract}


\ifCLASSOPTIONpeerreview
\begin{center} \bfseries EDICS Category: 3-BBND \end{center}
\IEEEpeerreviewmaketitle
\fi

\section{Introduction}
\IEEEPARstart{N}{onlinear} dynamic inversion (NDI) is increasingly becoming a popular control technique for modern high performance aerospace systems, primarily due to inherent automatic gain scheduling \cite{Oland2020,Tal2021}. Part of NDI which inverts control effector model is otherwise known as control allocation. Modern aircrafts are often designed to be overactuated with each control effector influencing multiple axes. The modular nature of NDI allows to re-assign/reconfigure the role of each control effector, according to requirements of operating flight regime, using only control allocation and without modification of baseline control law \cite{Yang2020,Shen2017}. Using a control allocation method the desired commands are distributed over the available control effector suite, such a way that the desired control effect is produced, along with fulfilling some additional objectives like minimization of deflections, drag and/or radar cross-section etc. In modern tailless configurations, such as innovative control effector concept, sometimes it becomes necessary to exploit the secondary axis yaw-power of elevons to provide artificial directional stability \cite{Niestroy2017,Dorsett1996}. Usually, yaw-power of an elevon is a highly nonlinear and asymmetric function of deflection. Other control devices, like spoiler-slot-deflectors and clamshell surfaces, also show nonlinear characteristics as well as control interactions \cite{Rajput2018}. Thus, a suitable nonlinear control allocation method is necessarily required to compute surface deflections that accurately produce the demanded control effect \cite{Benosman2009}.


A popular and computationally efficient framework for solving the nonlinear control allocation problem is known as the affine control allocation, in which the locally affine approximation of a nonlinear model of control surface effectiveness is computed at every sampling instant, and used as an input to a linear control allocation method \cite{Doman2002,Oppenheimer2011}. A discrete-time form of the affine control allocation that is suitable for implementation on digital computers is known as the incremental (or Frame-wise) control allocation \cite{Bordignon1996,Durham2017,Matamoros2018}. The advantage of the incremental control allocation is that it can also account for the actuator rate-limits, actuator dynamics and interactions between control effectors. Other approaches include sequential quadratic programming \cite{Johansen2004}, mixed-integer linear programming \cite{Bolender2004}, rule-based linear programming \cite{Bolender2005}, second-order cone programming \cite{Hu2018} and neural networks \cite{Yu2019,Khan2022} etc.

A nonlinear control allocation method requires an accurate model of control surface effectiveness. Traditionally, this model is obtained by approximating the aerodynamic (wind-tunnel or computational fluid dynamics (CFD) based) data of control surfaces using ordinary polynomials \cite{Bordignon1996}. The aerodynamic data normally depends on multiple variables, for which use of multidimensional polynomials can provide limited modeling accuracy. Moreover, high-degree polynomials exhibit Runge’s phenomenon leading to false non-monotonicity, which may render affine control allocation inapplicable \cite{Haerkegaard2003}. Recently, efforts have been made to improve the accuracy of polynomial-based onboard models using Simplex Spline functions \cite{Tol2013,Tol2014}. Simplex splines show higher approximation power than ordinary polynomials. Main disadvantages of spline-based effector-model are that it requires very large number of coefficients and, local effectiveness matrix is obtained through complicated transformation from Barycentric-coordinates to Cartesian-coordinates. These disadvantages increase the implementation complexity and add-up to the computational cost.

In this paper, a new nonlinear control allocation method is presented, which incorporates the control surface model based on a piecewise multi-linear representation (PMLR). The proposed PMLR is developed by combining a new generalization of Kronecker product for block matrices with the canonical piecewise linear representation of nonlinear functions \cite{Chua1988,Kahlert1990}. Since, aerodynamic data of an aircraft's control surfaces, obtained through either wind-tunnel testing or CFD simulations, is available at grid points and usually linearly interpolated in between, for the  nonlinear 6-DOF simulation of an aircraft, therefore it is inherently a piecewise multi-linear function of states and control surface deflections. The proposed formulation gives an exact model of such data and thus is capable of accurately modeling nonlinear control surface effectiveness over the entire flight envelope of an aircraft. The nonlinear 6-DoF simulation model of a miniature tailless flying wing aircraft is used in this work to evaluate the performance of proposed PMLR-based nonlinear control allocation. This aircraft has ten independently operating aerodynamic control surfaces with nonlinear moment versus deflection characteristics and control interactions. A modular control law, comprising the control allocation method and a baseline NDI controller, is designed for this aircraft to evaluate the control allocation performance.

The rest of paper is organized as follows: Sections 2 recapitulates the incremental nonlinear control allocation. In section 3, the proposed piecewise multi-linear representation is presented. In section 4, a PMLR-based nonlinear control law design is presented, which incorporates the proposed nonlinear control allocation method. In section 5, the proposed and polynomial based nonlinear control allocation methods are evaluated and compared using closed-loop 6-DOF simulation results. Section 6 ends this article with concluding remarks on the presented work. 

\section{Incremental Nonlinear Control Allocation}
Generally, a nonlinear control allocation method solves an underdetermined, typically constrained, system of equations representing a nonlinear mapping between control surface deflections $\delta\in\R{\kappa}$ and corresponding control effect, $T_\delta \in \R{d}$, where $\kappa
> d$. This nonlinear mapping is given as,
\begin{equation}\label{eq:04}
  T_\delta = g(x,\delta)
\end{equation}
\noindent
where $x\in\R{\sigma}$ is a vector of aerodynamic state-variables. Given the required control effect (the control demand), $T_{dem}\in\R{d}$, $\delta$ is sought such that
\begin{equation}\label{eq:06}
  g(x,\delta) = T_{dem}
\end{equation}
subject to
\begin{equation}\label{eq:07}
\begin{split}
   \delta_{min} \leq \delta \leq \delta_{max} \\
   |\dot{\delta}| \leq \dot{\delta}_{max}
\end{split}
\end{equation}
\noindent
where $\delta_{min}\in\R{\kappa}$ and $\delta_{max}\in\R{\kappa}$ are the vectors of upper and lower limits of deflections, respectively, $\dot{\delta}_{max}\in\R{\kappa}$ are control rate-limits.

With incremental re-formulation of above nonlinear control allocation problem \cite{Bordignon1996,Matamoros2018}, instead of solving for complete magnitude of deflection, the solution for increment in previous deflection is sought at every sampling instant. Thus, one requires to find incremental deflection, $\Delta\delta$, such that
\begin{eqnarray}
  G \Delta \delta = \Delta T_{dem} \label{eq:15}\\
  \Delta\underline{\delta} \leq \Delta\delta \leq \Delta\bar{\delta} \label{eq:16}
\end{eqnarray}
\noindent
where, $G\in\R{d\times\kappa}$ is the control effectiveness matrix
\begin{equation}\label{eq:09}
  G = \left(\frac{\partial g}{\partial \delta}\right)_{\left(x_0,\delta_0\right)}
\end{equation}
and,
\begin{equation}\label{eq:17}
  \Delta T_{dem} = T_{dem} - g(x_0,\delta_0)
\end{equation}
Here $g(x_0,\delta_0)$ is the onboard estimation of control-effect produced by previous deflection, $\delta_0$. The incremental limits, $\Delta\underline{\delta}\in\R{\kappa}$ and $\Delta\bar{\delta}\in\R{\kappa}$, are specified as
\begin{equation}\label{eq:18}
\begin{split}
\Delta \bar{\delta} &= \min(\delta_{max}-\delta_0, \dot{\delta}_{max}\Delta t)  \\
\Delta \underline{\delta} &= \max(\delta_{min}-\delta_0,- \dot{\delta}_{max}\Delta t)
\end{split}
\end{equation}
The vector of total actuator commands, $\delta$, is then computed as,
\begin{equation}\label{eq:19}
  \delta = \delta_0 + \Delta\delta
\end{equation}
Resulting affine control allocation problem of Eq. \eqref{eq:15}-\eqref{eq:16} is then be solved by any of the several available efficient and well-tested linear control allocation solvers \cite{Oppenheimer2011,Bodson2002}.

The affine/incremental framework of nonlinear control allocation requires an updated local control-effectiveness matrix, $G$, at every sampling instant, in order to compute control surface deflections. Thus, accuracy of nonlinear control allocation greatly depends upon how good the matrix, $G$, locally approximates the nonlinear effectiveness of a control surface suit, given as $g(x,\delta)$. Thus, the estimation of matrix, $G$, requires an onboard implementation of the control surface model, $g(x,\delta)$. This model is also required for the computation of incremental control demand, $\Delta T_{dem}$, at every sampling instant.

\section{Piecewise Multi-Linear Representation}
In this section, we develop a new formulation to model multi-variate piecewise multi-linear functions, which is used subsequently to develop control surface models for use with incremental control allocation. First, a new generalization of Kronecker product for block matrices is presented. Such a generalizations are often termed as the block Kronecker products (see for example \cite{Tracy1972,Hyland1989,Koning1991}). This block Kronecker product is then combined with the canonical piecewise linear representation (\cite{Chua1988,Kahlert1990}) to form a piecewise multi-linear representation.

\begin{definition}[Block Kronecker Product]\label{def:BKP} 
Let $\A$ is a $m \times n$ matrix and $\B$ is a $p \times q$ matrix, where $n = \kappa p$ for some positive integer $\kappa$. Then the Block Kronecker Product is defined in two steps:
\begin{enumerate}
  \item  Partition $\A$ into $\kappa$ blocks $A_i$, each of size $m \times p$ i.e.
  \begin{equation}\label{eq:43A}
    \A = \begin{bmatrix} A_{1} & \cdots & A_{\kappa} \end{bmatrix}
  \end{equation}
  \item Perform the product operation as follows,
  \begin{equation}\label{eq:43}
    \A \LBKP \B = \begin{bmatrix} A_{1}\B & \cdots & A_{\kappa}\B \end{bmatrix}
  \end{equation}
  where `$\LBKP$' is defined as the Block Kronecker product operator.
\end{enumerate}
\end{definition}


\begin{remark}
It must be noted that the condition of existence of the Block Kronecker Product ($\A \LBKP \B$), is that the number of columns of $\A$ must be integer multiple of the number of rows of $\B$. In case if both are equal, then this product reduces to standard matrix product.
\end{remark}

Now let us present some important properties of Block Kronecker Product.
\begin{enumerate}
  \item \textbf{\textit Homogeneity}: For any scalar $\lambda$, and matrices $\A$ and $\B$ of appropriate sizes we have
\begin{equation}\label{eq:BKP_Prop1}
\begin{split}
   \A \LBKP (\lambda \B)  &= \lambda(\A \LBKP \B) \\
   (\lambda \A) \LBKP \B  &= \lambda(\A \LBKP \B)
\end{split}
\end{equation}
  \item \textbf{\textit Distributivity}: For any matrices $\A$, $\B$, $\A_1$, $\A_2$, $\B_1$, and $\B_2$ of appropriate sizes we have,
\begin{equation}\label{eq:BKP_Prop2}
\begin{split}
   \A \LBKP (\B_1+\B_2) = (\A \LBKP \B_1) + (\A \LBKP \B_2) \\
   (\A_1+\A_2) \LBKP \B = (\A_1 \LBKP \B) + (\A_2 \LBKP \B)
\end{split}
\end{equation}
  \item \textbf{\textit Associativity}: For any matrices $\A$, $\B$, and  $\C$ of appropriate sizes we have,
\begin{equation}\label{eq:BKP_Prop3}
   \A \LBKP (\B \LBKP \C) = (\A \LBKP \B) \LBKP \C
\end{equation}
  \item \textbf{\textit Inverse}: For any invertible matrix $\A$ with inverse $\A^{-1}$,
\begin{equation}\label{eq:BKP_Prop4}
   \A \LBKP \A^{-1} = \A^{-1} \LBKP \A = \mathbb{I}
\end{equation}
\end{enumerate}

The proof of these properties directly follows from the definition \eqref{eq:43}. Note that \emph{linearity} of the operator is an immediate consequence of \emph{homogeneity} and \emph{distributivity}.

Using definition \ref{def:BKP} now we present piecewise multi-linear  representation in the theorem below.

\begin{theorem}[Piecewise Multi-Linear Representation]\label{THRM:1}
Consider a vector-valued piecewise multi-linear function $g:\R{k}\mapsto\R{m}$, whose values are given
along a rectilinear grid $(z_1,z_2,\cdots,z_k)$ at $[\mu_1^{(1)},\cdots, \mu_1^{(L_1)}]\times[\mu_2^{(1)}, \cdots,\mu_2^{(L_2)}]\times\cdots\times [\mu_k^{(1)},\cdots,\mu_k^{(L_k)}]$. Then the $g(z_1,z_2,\cdots,z_k)$ can be compactly expressed in piecewise multi-linear representation as,
\begin{equation}\label{eq:45}
  g = \left( \left(\left( \Gamma \LBKP\hat{z}_k \right)\LBKP\hat{z}_{k-1} \cdots \right)\LBKP\hat{z}_2 \right)\hat{z}_1
\end{equation}
where $\Gamma \in \R{m\times\prod_{j=1}^{k}L_j}$, and
\begin{equation}\label{eq:46}
  \hat{z}_j = \begin{bmatrix} 1 \\ z_j \\ \left|z_j - \mu_j^{(2)}\right| \\ \vdots \\ \left|z_j - \mu_j^{(L_j-1)}\right| \end{bmatrix},\qquad \forall\,j \in [1,k]
\end{equation}
\end{theorem}
\begin{proof}
Since any single variable piecewise linear function can be written in canonical form \cite{Chua1988,Kahlert1990} as
\begin{equation}\label{eq:a}
  f(z) = \gamma \hat{z}
\end{equation}
where $\gamma = [\gamma^{(1)},\gamma^{(2)},\cdots,\gamma^{(L)}]$ is a matrix of coefficients, then we can write the multivariate function $g(z_1,z_2,\cdots,z_k)$ as
\begin{equation}\label{eq:b}
  g = \Gamma_1(z_2,\cdots,z_k) \hat{z}_1
\end{equation}
where
\begin{equation}\label{eq:c}
\begin{split}
\Gamma_1 &= \left[\gamma_1^{(1)}(z_2,\cdots,z_k),\cdots,\gamma_1^{(L_1)}(z_2,\cdots,z_k)\right] \\
&= \left[\gamma_2^{(1)}(z_3,\cdots,z_k)\hat{z}_2,\cdots,\gamma_2^{(L_1)}(z_3,\cdots,z_k)\hat{z}_2\right] \\
&= \Gamma_2 \LBKP \hat{z}_2
\end{split}
\end{equation}
Similarly, we can write,
\begin{equation}\label{eq:d}
  \Gamma_{j} = \Gamma_{j+1} \LBKP \hat{z}_{j+1}, \quad \forall j\in[1,k-1]
\end{equation}
Now, continuing the expansion of Eq. \eqref{eq:b} as described by Eq. \eqref{eq:d} for $k-1$ times (till $\hat{z}_k$) results in Eq. \eqref{eq:45}, where $\Gamma = \Gamma_k$, which completes the proof.
\end{proof}

\begin{corollary}[Partial Differentiation]\label{COROL:1}
From Eqs. \eqref{eq:45} and \eqref{eq:46}, it directly follows that
\begin{equation}\label{eq:Diff}
\frac{\partial g}{\partial \hat{z}_j} = \left(\left(\left( \Gamma \LBKP\hat{z}_k \cdots \right)\LBKP\frac{\partial\hat{z}_j}{\partial z_j}\right)\LBKP \hat{z}_{j-1} \cdots \right)\hat{z}_1
\end{equation}
where,
\[
\frac{\partial\hat{z}_j}{\partial z_j} = \begin{bmatrix} 0 \\ 1 \\ \sign\left(z_j - \mu_j^{(2)}\right) \\ \vdots \\
 \sign\left(z_j - \mu_j^{(L_j-1)}\right) \end{bmatrix},\qquad \forall\,j\in [1,k]
\]
\end{corollary}
\begin{proof}
Proof follows directly from linearity of Block Kronecker Product which is evident from its properties Eq. \eqref{eq:BKP_Prop1} and \eqref{eq:BKP_Prop2}.
\end{proof}

Next, a method is presented to compute $\Gamma$ from given values of the function $g(z)$ along the $k$-dimensional rectilinear grid points. Before proceeding further, let us define a matrix reshaping transformation.
\begin{definition}[Reshaping Transformation]\label{def:RSTran}
The reshaping transformation $\mathcal{T}_\lambda: \R{m\times n} \mapsto \R{\lambda\times\kappa}$, such that $mn = \lambda\kappa$,  is defined as,
\begin{equation}\label{eq:RSTrandef}
\TR{\lambda}{\A} = \left(\mathrm{vec}(\mathbb{I}_\kappa)^\top \otimes \mathbb{I}_\lambda \right) \LBKP  \mathrm{vec}(\A)
\end{equation}
where $\mathrm{vec}(\cdot)$ denotes standard vectorization operation,  $\mathbb{I}$ denotes identity matrix, and  `$\otimes$' represents standard Kronecker product.
\end{definition}

Now consider a matrix $\A \in\R{m \times n}$, a set vectors $x_i\in\R{L}$ for all $i\in[1,L]$, and some integer $\lambda$ such that $mn = \lambda\kappa$ for some positive integer $\kappa$. Let $B_i = \A \LBKP x_i$. Then if we define $\hat{X} = \begin{bmatrix}x_1 & \cdots & x_L \end{bmatrix}$, and $\B^\top = \begin{bmatrix}B_1^\top & \cdots & B_L^\top \end{bmatrix}$, then following identities holds,
\begin{equation}\label{eq:RSTran01}
\A = \TR{m}{\TR{\lambda}{\A}}
\end{equation}
\begin{equation}\label{eq:RSTran02}
\B = \TR{mL}{\A\LBKP\hat{X}}
\end{equation}
The proof of these relations directly follows from definitions \ref{def:BKP} and \ref{def:RSTran}, and properties \eqref{eq:BKP_Prop1}-\eqref{eq:BKP_Prop4} and is omitted here for brevity. Next theorem presents the formulation to compute $\Gamma$ from a given dataset.

\begin{theorem}[PMLR Fitting]\label{THRM:2}
Given a $k$-dimensional data $Y_i\in\R{L_1\times \cdots \times L_k}$  for each output $i\in[1,m]$, at a rectilinear grid defined as $[\mu_1^{(1)},\cdots, \mu_1^{(L_1)}]\times[\mu_2^{(1)}, \cdots,\mu_2^{(L_2)}]\times\cdots\times [\mu_k^{(1)}, \cdots,\mu_k^{(L_k)}]$. Then $\Gamma$ in Eq. \eqref{eq:45} can be computed as follows,
\begin{equation}\label{eq:TH02}
\Gamma = \begin{bmatrix} Q_k^1 \\ \vdots \\ Q_k^m \end{bmatrix}
\end{equation}
where,
\begin{equation*}
\left.\begin{split}
Q_0^i &= \mathrm{vec}(\hat{Y_i}) \\
Q_j^i &= \TR{\lambda_j}{Q^i_{j-1}} \LBKP \hat{Z}_{j}^{-1}, \quad \forall\,j\in[1,k]\\
\end{split}\right\},\quad\forall\,i\in[1,m]
\end{equation*}
where $\hat{Y}_i$ is obtained by flipping dimensions of $k$-dimensional array $Y_i$ for each $i\in[1,m]$, and
\begin{equation*}
\left.\begin{split}
\hat{Z}_j &= \begin{bmatrix} \left.\hat{z}_j\right|_{z_j = \mu_j^{(1)}} & \cdots &  \left.\hat{z}_j\right|_{z_j = \mu_j^{(L_j)}} \end{bmatrix}\\
\lambda_j &= \prod_{i=j+1}^{k}L_k
\end{split}\right\},\quad\forall\,j\in[1,k]
\end{equation*}
\end{theorem}

\begin{proof}
Due to structure of $\Gamma$ in \eqref{eq:TH02}, it is sufficient to show only scalar-valued case ($m=1$), since generalization to vector-valued is straight forward, So we will only show the scalar-valued case. Let $y_{(i_1,i_2,\cdots,i_k)}$ represents the output data at $[\mu_1^{(i_1)},\mu_2^{(i_2)},\cdots, \mu_k^{(i_k)}]$. Then using equations \eqref{eq:b} and \eqref{eq:d} we can write,
\begin{equation}\label{eq:th2p_1}
\begin{split}
y_{(i_1,i_2,\cdots,i_k)} &= \Gamma_1^{(i_2,\cdots,i_k)} \LBKP \hat{z}_1^{(i_1)} \\
\Gamma_{j-1}^{(i_{j},\cdots,i_k)} &= \Gamma_{j}^{(i_{j+1},\cdots,i_k)} \LBKP \hat{z}_j^{(i_j)},\quad\forall\;j\in[2,k-1] \\
\Gamma_{k-1}^{(i_k)} &= \Gamma \LBKP \hat{z}_k^{(i_k)}
\end{split}
\end{equation}
where $\hat{z}_j^{(i_j)} = \left.\hat{z}_j\right|_{z_j = \mu_j^{(i_j)}}$,Now lets denote,
\begin{equation}
Q_0 = \begin{bmatrix}
y_{(1,1,\cdots,1)}   \\ \vdots \\ y_{(1,1,\cdots,L_k)} \\ \vdots \\
y_{(1,1,\cdots,L_{k-1},1)} \\ \vdots \\
y_{(1,1,\cdots,L_{k-1},L_k)} \\ \vdots \\ y_{(L_1,L_2,\cdots,L_{k-1},L_k)} \end{bmatrix}
\end{equation}
Since $y_{(i_1,i_2,\cdots,i_k)}$ are elements of given dataset $Y$, here it must be noted that $Q_0 = \mathrm{vec}(\hat{Y})$. Also let
\begin{equation}
Q_j = \begin{bmatrix}
\Gamma_j^{(1,1,\cdots,1)}   \\ \vdots \\ \Gamma_j^{(1,1,\cdots,L_k)} \\ \vdots \\
\Gamma_j^{(1,1,\cdots,L_{k-1},1)} \\ \vdots \\
\Gamma_j^{(1,1,\cdots,L_{k-1},L_k)} \\ \vdots \\ \Gamma_j^{(L_{j+1},L_{j+2},\cdots,L_{k-1},L_k)} \end{bmatrix}, \quad \forall\;j\in[1,k-1]
\end{equation}
Also since $m=1$, so $\Gamma = Q_k$. Then we can write \eqref{eq:th2p_1}, by using \eqref{eq:RSTran02} as follows
\begin{equation}\label{eq:th2p_2}
\begin{split}
Q_{j-1} &= \TR{L_j}{Q_j \LBKP \hat{Z}_j},\qquad\forall\;j\in[1,k] \\
\end{split}
\end{equation}
Now since $Q_j\in\R{\lambda_j\times \bar{\lambda}_j}$, where $\lambda_j = \prod_{i=j+1}^{k}L_i$ and $\bar{\lambda}_j = \prod_{i=1}^{j}L_i$, for each $j\in[0,k]$. Thus  by using \eqref{eq:RSTran01} we get
\begin{equation}\label{eq:th2p_3}
\begin{split}
Q_{j} &= \TR{\lambda_j}{Q_{j-1}} \LBKP \hat{Z}_j^{-1},\qquad\forall\;j\in[1,k] \\
\end{split}
\end{equation}
which concludes the proof.
\end{proof}

\begin{lemma}\label{LEM:01}
For any vectors $x \in \R{n}$ and $y \in \R{m}$ and any matrix $\A \in \R{q \times mn}$, the following relation holds,
\begin{equation}\label{eq:lem1}
  (\A \LBKP y) x = \A (x \otimes y)
\end{equation}%
\end{lemma}
\begin{proof}
Using Eq. \eqref{eq:43} we can write
\begin{equation*}
\begin{split}
(\A \LBKP y) x &= \begin{bmatrix} A_1 y & A_2 y & \cdots & A_n y \end{bmatrix} x \\
&= \begin{bmatrix} A_1 & A_2 & \cdots & A_n \end{bmatrix} \begin{bmatrix}  x^{(1)} y \\ x^{(2)} y \\ \vdots \\ x^{(n)} y \end{bmatrix} = \A (x \otimes y)
\end{split}
\end{equation*}
where $x^{(i)}$ represents $i$th element of vector $x$.
\end{proof}

Next, an alternative formulation is presented which uses standard Kronecker product operation to represents the nested formulation of Eq. \eqref{eq:45}.

\begin{theorem}[Alternative Representation]\label{THRM:3}
The piecewise multi-linear function $g$  (given by Eq. \eqref{eq:45}) can be equivalently expressed as follows,
\begin{equation}\label{eq:BKP_Kron1}
  g = \Gamma \bigotimes_{i=1}^{k} \hat{z}_i
\end{equation}
where $\bigotimes_{i=1}^{k} \hat{z}_i$ is the compact representation of the successive standard Kronecker product operation $\hat{z}_1 \otimes \hat{z}_2 \otimes \cdots \otimes \hat{z}_k$.
\end{theorem}
\begin{proof}
First consider the two-dimensional case ($k=2$) and using Lemma \ref{LEM:01}, we have
\begin{equation}
  (\Gamma_2 \LBKP \hat{z}_2)\hat{z}_1 = \Gamma_2 (\hat{z}_1 \otimes \hat{z}_2)
\end{equation}
Thus Eq. \eqref{eq:BKP_Kron1} is true for $k=2$. Now, assume that it is also true for $k=j$, so
\begin{equation}
\left(\left( \Gamma_j \LBKP\hat{z}_j \right)\LBKP\hat{z}_{j-1} \cdots \right)\LBKP\hat{z}_1 = \Gamma_j \bigotimes_{i=1}^{j} \hat{z}_i
\end{equation}
Now, from Eq. \eqref{eq:d},
\begin{equation}
\left(\left( \Gamma_{j+1} \LBKP\hat{z}_{j+1} \right)\LBKP\hat{z}_{j} \cdots \right)\LBKP\hat{z}_1 = \left( \Gamma_{j+1} \LBKP\hat{z}_{j+1} \right) \bigotimes_{i=1}^{j} \hat{z}_i
\end{equation}
applying Lemma \ref{LEM:01} again yields,
\begin{equation}
\left(\left( \Gamma_{j+1} \LBKP\hat{z}_{j+1} \right)\LBKP\hat{z}_{j} \cdots \right)\LBKP\hat{z}_1 =  \Gamma_{j+1} \bigotimes_{i=1}^{j+1} \hat{z}_i
\end{equation}
Hence, by the principal of mathematical induction Eq. \eqref{eq:BKP_Kron1} is true.
\end{proof}


\begin{corollary}\label{COROL:2}
From Eq. \eqref{eq:BKP_Kron1}, it follows that
\begin{equation}\label{eq:BKP_Kron2}
 \frac{\partial g}{\partial \hat{z}_j} = \Gamma \left[\left(\bigotimes_{i=1}^{j-1} \hat{z}_i\right) \otimes \frac{\partial\hat{z}_j}{\partial z_j} \otimes \left(\bigotimes_{i=j+1}^{k} \hat{z}_i\right)\right]
\end{equation}
\end{corollary}
\begin{proof}
Its proof is direct consequence of Theorems \ref{THRM:1} and \ref{THRM:3} and Corollary \ref{COROL:1}.
\end{proof}

The advantage of alternative representation Eq. \eqref{eq:BKP_Kron1} is two-fold. First, from on-board implementation standpoint, alternative representation provides a straight forward formulation as compared to Eq. \eqref{eq:45}. Secondly, a linear regression scheme can also be formulated directly from Eq. \eqref{eq:BKP_Kron1} for computation of the coefficient matrix $\Gamma$. Given the data at each grid point, i.e. a total of $n=\prod_{j=1}^{k}L_j$ grid points. Let $x_1,x_2,\cdots,x_n \in \R{n}$ represent values of $\bigotimes_{i=1}^{k} \hat{z}_i$, evaluated using Eq. \eqref{eq:46} at each grid point, and let $y_1,y_2,\cdots,y_n\in\R{m}$ represent corresponding output vectors. Now defining the observation matrix $Y = [y_1,y_2,\cdots,y_n]$, and the regressor matrix $X = [x_1,x_2,\cdots,x_n]$, we can write Eq. \eqref{eq:BKP_Kron1} collectively for all data points in following compact form,

\begin{equation}\label{eq:SolveGamma}
  Y = \Gamma X
\end{equation}

So $\Gamma$ can be found by solving linear system Eq. \eqref{eq:SolveGamma}, as $\Gamma =  Y X^{-1}$. It must be noted that simple inverse is required to compute $\Gamma$, which gives an extra evidence of the fact that the multi-linear representation  of Eq. \eqref{eq:45} is an exact fit of a multivariate piecewise multi-linear data.

For the application of nonlinear control allocation, where number of independent variable are usually less than or equal to five, Eq. \eqref{eq:SolveGamma} can be used to directly compute $\Gamma$. However, for the applications having large number of independent variables, solution of Eq. \eqref{eq:SolveGamma}, may out-run the computational resources of an average PC or laptop. Therefore, the iterative algorithm (Theorem \ref{THRM:2}) is recommended. However, for onboard implementation of PMLR, alternative representation (Theorem \ref{THRM:3}) is recommended due to its distinctive features.

\section{PMLR Based Nonlinear Control Law}
\begin{figure}
  \centering
  \includegraphics[height=0.5\linewidth]{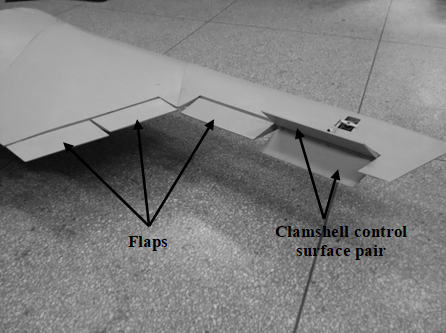}
  \caption{Aircraft Control Surface Assembly}\label{fig:XQ6B}
\end{figure}

\begin{figure}
  \centering
  \includegraphics[height=0.5\linewidth]{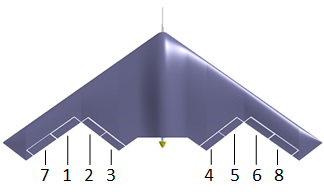}
  \caption{Numbering scheme of control surfaces}\label{fig:CSNumbering}
\end{figure}

\begin{figure*}
  \centering
  \includegraphics[width=0.85\linewidth]{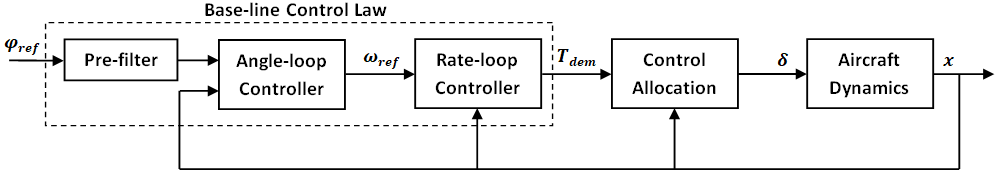}
  \caption{Schematic of Complete Control Law}\label{fig:ContLaw}
\end{figure*}

In this section a PMLR based nonlinear control law is presented for a miniature tailless flying wing aircraft. The detailed specification of the aircraft can be found in \cite{Qu2017}. The aircraft has six flaps (control surfaces at the trailing-edge of the wing) and, two (left/right) pairs of clamshell surfaces (see Fig. \ref{fig:XQ6B}). The numbering scheme of control surfaces is shown in Fig. \ref{fig:CSNumbering}. The control law consists of the baseline control law and the incremental control allocation, as shown in Fig. \ref{fig:ContLaw}. The baseline control law is based on nonlinear dynamic inversion (NDI), which generates control demand in terms of a moment vector, $T_{dem}$ \cite{Enns1994,Lane1988}. Control demand is then fed to the incremental control allocation algorithm to compute control commands for each control surface. Incremental control allocation algorithm is based on the Redistributed Pseudo-Inverse (RPI) method \cite{Oppenheimer2011}, which incorporates either PMLR- or polynomial-based control surface model. The control system is purposefully made sensitive to control surface model inaccuracies by not including integral-action in the baseline control law so that the effects of control allocation errors could be made more noticeable.
\subsection{Aircraft Model}
Aerodynamic moment-model of the aircraft is given as under,
\begin{equation}\label{eq:AeroModel}
T = q_\infty S L_{ref}\left(C_A + \dfrac{1}{2V} L_{ref} C_\omega \omega + C_\delta\right)
\end{equation}
\noindent
where $T\in\R{3}$ represents net moments, $q_\infty$ is free-stream dynamic pressure, $V$ is aircraft's true velocity, $S$ is reference area, $L_{ref} = \mathrm{diag}(b,\bar{c},b)$ with $\bar{c}$ and $b$ being mean aerodynamic chord and wing span, respectively, $\omega\in\R{3}$ is body-axes angular rates, $C_A\in\R{3}$ represents aerodynamic moment coefficients at zero deflections, $C_\omega\in\R{3\times 3}$ is matrix of damping derivatives, and $C_\delta \in\R{3}$ is vector of increment in aerodynamic moment coefficients due control surface deflections which is given as
\begin{equation}\label{eq:ContrSurfModel}
\begin{split}
C_\delta =& \sum_{i=1}^{6}\VEC{\Delta C_l(\alpha,\delta_i)}{\Delta C_m(\alpha,\delta_i)}{\Delta C_n(\alpha,\delta_i)} \\
&{}+ \sum_{i=7}^{8}\VEC{\Delta C_l(\alpha,\beta,\delta_{iU},\delta_{iL})}{\Delta C_m(\alpha,\beta,\delta_{iU},\delta_{iL})}{\Delta C_n(\alpha,\beta,\delta_{iU},\delta_{iL})}
\end{split}
\end{equation}
where $\Delta C_l$, $\Delta C_m$, and $\Delta C_n$ are change in roll, pitch and yaw moment coefficients, respectively, due to control surface deflections, which are also function of the angle of attack $\alpha$ and the side-slip angle $\beta$.

All moment contributions due to deflection of flaps ($\delta_i$) are two-dimensional functions.
Flaps are known to have highly nonlinear and asymmetric yaw moment effectiveness about zero-deflection. Moreover, they show significant nonlinearities in pitch and roll moments for large deflection angles \cite{Stenfelt2009,Stenfelt2010}. Moment contributions of each clamshell pair are four-dimensional functions. The deflections of lower and upper clamshell surfaces are referred to as, $\delta_{iL}$ and $\delta_{iU}$, respectively. Traditionally, clamshell pair is used as a pure yaw control device by deflecting the lower and upper surfaces by an equal amount, thus, effectively nullifying the effectiveness in roll and pitch axes and maximizing the effectiveness in yaw axis. Apart from the traditional usage, it is also possible to use each surface in the clamshell pair as a separate control device affecting multiple axes of rotation. Such an innovative usage certainly increases the redundancy of control surface suit. However, it requires modeling the clamshell pair data as an additively non-separable function with respect to lower and upper surfaces, in order to account for possible control interaction effects between these surfaces \cite{Rajput2018,Oppenheimer2007}. The clamshell surface pair shows fairly linear characteristics in roll and pitch axes, but, significantly nonlinear characteristics in yaw axis \cite{Qu2017,Rajput2014}.

\subsection{PMLR based Incremental Control Allocation}
From the aerodynamic model given in Eq. \eqref{eq:AeroModel}, the control surface effectiveness can be written as,
\begin{equation}\label{eq:05}
   \begin{split}
   g(x,\delta) \triangleq T_\delta &= q_\infty S L_{ref} C_\delta
   \end{split}
\end{equation}
The PMLR-based control surface model of the aircraft is given as
\begin{equation}\label{eq:48}
\begin{split}
C_\delta =& \sum_{i=1}^{6}\left(\Gamma_{i}\LBKP\hat{\alpha}\right)\hat{\delta}_i \\
&{}+ \sum_{i=7}^{8}\left(\left(\left(\Gamma_{i}\LBKP\hat{\alpha}\right)\LBKP\hat{\beta}\right)\LBKP \hat{\delta}_{iL}\right)\hat{\delta}_{iU}
\end{split}
\end{equation}
Partial derivatives of above control surface model along all control variables are required in order to compute elements of the control effectiveness matrix, $G$  ($= q_\infty S L_{ref} M$), where $M\triangleq\partial C_\delta/\partial\delta$ is given as
\begin{equation}\label{eq:49}
M = \begin{bmatrix} \s_{\delta_1} & \cdots & \s_{\delta_6} & \s_{\delta_{7L}} & \s_{\delta_{7U}} & \s_{\delta_{8L}} & \s_{\delta_{8U}} \end{bmatrix}
\end{equation}

Elements of control effectiveness matrix are computed, at every sampling instant, from PMLR-based control surface model as follows:
\begin{equation}\label{eq:50}
\begin{split}
\s_{\delta_i} &= \left(\Gamma_{i}\LBKP \hat{\alpha}\right) \frac{\mathrm{d} \hat{\delta}_i}{\mathrm{d}\delta_i},\quad i=1,2,\cdots,6 \\
\s_{\delta_{iU}} &=  \left(\left(\left(\Gamma_{i}\LBKP \hat{\alpha}\right)\LBKP\hat{\beta}\right)\LBKP \hat{\delta}_{iL}\right) \frac{\mathrm{d} \hat{\delta}_{iU}}{\mathrm{d}\delta_{iU}},\quad i=7,8 \\
\s_{\delta_{iL}} &=  \left(\left(\left(\Gamma_{i}\LBKP \hat{\alpha}\right)\LBKP\hat{\beta}\right)\LBKP\frac{\mathrm{d} \hat{\delta}_{iL}}{\mathrm{d}\delta_{iL}} \right) \hat{\delta}_{iU} ,\quad i=7,8
\end{split}
\end{equation}

Now, it is straightforward to compute $g(x_0,\delta_0)$ from Eqs. \eqref{eq:05}-\eqref{eq:48}. Computed control effectiveness matrix and $g(x_0,\delta_0)$ are then used in Eqs. \eqref{eq:15}-\eqref{eq:17} to form the incremental control allocation problem, which is then solved by using the RPI method as follows:
\begin{equation}\label{eq:53}
  \Delta \delta  = \delta_p + G^{\dag}\left[\Delta T_{dem} - G\delta_p\right]
\end{equation}
\noindent
where
\begin{equation}\label{eq:54}
  G^{\dag} = W^{-1}G^{T}(GW^{-1}G^{T})^{-1}
\end{equation}
\noindent
is the weighted pseudoinverse of matrix $G$, $\delta_p \in \R{10}$ is a preference vector and, $W\in\R{10\times 10}$ is a weighting matrix. The constraints of actuator position and rate are incorporated through the redistribution process as given in \cite{Oppenheimer2011}.

Various components of control surface aerodynamic data of the aircraft, as given in Eq. \eqref{eq:ContrSurfModel}, are modeled using traditional least square fittings of multivariable polynomials and also using the proposed PMLR formulation. Due to aircraft symmetry, it is sufficient to produce models of starboard control surfaces only. Both effector models are validated by passing 10,000 samples of randomly generated input vector containing surface deflections and aerodynamic angles through the actual linearly interpolated lookup tables of aerodynamic data and the control surface models. It was observed that, for all polynomial models, relative RMS errors are less than $9\%$ and maximum errors are less than $0.0015$.
Small values of errors indicate good accuracy of resulting polynomial models. Since, PMLR-based effector model is an exact representation of aerodynamic data, its error is zero. Such an accurate modeling, however, comes at the cost of relatively larger number of coefficients than the polynomial counterparts. Table \ref{tab:Comparison} gives a comparison of number of coefficients for each case, which shows that number of coefficients for PMLR model are moderately higher but acceptable for onborad implementation.

\renewcommand\arraystretch{1.3}
\begin{table}
\centering
\caption{No. of Coefficients}
\label{tab:Comparison}
\arrayrulecolor{black}
\begin{tabular}{ccc}
\toprule
Component & Polynomial & PMLR \\ \cmidrule(rl){1-3}
$\Delta C_l (\alpha,\delta_4)$ & 12 & 31 \\
$\Delta C_l (\alpha,\delta_5)$ & 20 & 31\\
$\Delta C_l (\alpha,\delta_6)$ & 16 & 31\\
$\Delta C_l (\alpha,\beta,\delta_{8L},\delta_{8U})$ & 108 & 385 \\ \cmidrule(rl){1-3}
$\Delta C_m (\alpha,\delta_4)$ & 12 & 31 \\
$\Delta C_m (\alpha,\delta_5)$ & 16 & 31\\
$\Delta C_m (\alpha,\delta_6)$ & 12 & 31\\
$\Delta C_m (\alpha,\beta,\delta_{8L},\delta_{8U})$ & 240 & 385\\ \cmidrule(rl){1-3}
$\Delta C_n (\alpha,\delta_4)$ & 20 & 31\\
$\Delta C_n (\alpha,\delta_5)$ & 20 & 31\\
$\Delta C_n (\alpha,\delta_6)$ & 16 & 31\\
$\Delta C_n (\alpha,\beta,\delta_{8L},\delta_{8U})$ & 192 & 385\\
\bottomrule
\end{tabular}
\arrayrulecolor{black}
\end{table}
\renewcommand\arraystretch{1.0}

\subsection{NDI Control Law}
The rate-loop NDI control law is written in terms of vector of demanded control moments, $T_{dem}$, which is translated into control surface deflections using a control allocation method. Consider the aircraft’s moment equations in vector form,
\begin{equation}\label{eq:55}
  \dot{\omega} = I^{-1}\left( T_a - \omega\times I \omega \right) + I^{-1}T_\delta
\end{equation}
\noindent
where $I\in\R{3\times 3}$ is the inertia matrix. $T_a\in\R{3}$  is the vector of aerodynamic moments due to airframe, which is given as follows:
\begin{equation}\label{eq:56}
  T_a = q_\infty S L_{ref} \left(C_A + \dfrac{1}{2V} L_{ref} C_\omega \omega\right)
\end{equation}
\noindent
where $x\in\R{6}$ is a vector of aircraft state-variables. $T_\delta\in\R{3}$ is the vector of moments due to control effector deflections. Now, let $\dot{\omega}_{des}\in\R{3}$ be the vector of desired angular body accelerations, then to achieve the control objective $\dot{\omega} = \dot{\omega}_{des}$, the inversion law is given as,

\begin{equation}\label{eq:57}
  T_{dem} = T_\delta = I\dot{\omega}_{des} - T_a + \omega\times I \omega
\end{equation}

In order to implement above control law, an onboard model of multivariate function ($T_a$) is required. This model is also implemented using PMLR. Full-state feedback-linearization of the nonlinear system of Eq. \eqref{eq:55} by the inversion law of Eq. \eqref{eq:57} results in a system of three integrators, $\dot{\omega} = \dot{\omega}_{des}$. The time-constant of each integrator channel can be set as desired by a linear controller. Thus, the complete rate-loop control law can be written as
\begin{equation}\label{eq:58}
  T_{dem} = I K_\omega (\omega - \omega_{ref}) -  T_a + \omega\times I \omega
\end{equation}
\noindent
where $K_\omega = \mathrm{diag}(k_p,k_q,k_r)$. The tuned gains of linear part of the inner-loop controller are listed in Table \ref{tab:LoopGains}.


For the angle-loop control law design, the aircraft dynamics with rate-loop closed is taken as the plant, whose input is a vector of reference angular rates, $\omega_{ref}$. The angle-loop controlled variables are roll angle $\phi$, pitch angle $\theta$, and sideslip angle $\beta$. The NDI control law of angle-loop, in terms of reference angular rates, is derived as follows.

Assuming small aerodynamic angles ($\sin\alpha \approx \sin\beta \approx 0$ and $\cos\alpha \approx 1$), the kinematic equations can be written as,
\begin{equation}\label{eq:59}
    \dot{\Phi} = \Lambda \omega + f_\Phi(x)
\end{equation}
\noindent
where $\Phi = \left[\phi,\theta,\beta\right]^\top$ is attitude angles, $\Lambda = \mathrm{diag}(1,\cos\phi,-1)$, and

\[
f_\Phi(x) = \begin{bmatrix} q\sin\phi\tan\theta + r\cos\phi\tan\theta \\ - r\sin\phi \\ \frac{1}{V}\bar{g}_0\cos\theta\sin\phi \end{bmatrix}
\]

Let $\dot{\Phi}_{des} = \left[\dot{\phi}_{des},\dot{\theta}_{des},\dot{\beta}_{des}\right]^\top$ be the desired attitude rate, then, to make $\dot{\Phi} = \dot{\Phi}_{des}$ the NDI control law in terms of reference body-axis roll-rate is obtained as

\begin{equation}\label{eq:60}
  \omega_{ref} = \omega = \Lambda^{-1}\left(\dot{\Phi}_{des} - f_\Phi(x)\right)
\end{equation}

As a result of above feedback-linearization, system of nonlinear state-equations is reduced to system of three integrators, $\dot{\Phi}=\dot{\Phi}_{des}$. The outer-loop controlled variables are then controlled by using linear proportional controller. Thus, the complete angle-loop control law can be written as
\begin{equation}\label{eq:67}
  \omega_{ref} = \Lambda^{-1}\left(K_\Phi(\Phi - \Phi_{ref}) - f_\Phi(x)\right)
\end{equation}
\noindent
where the vector of reference input is $\Phi_{ref}=\HVEC{\phi_{ref}}{\theta_{ref}}{\beta_{ref}}^T$ and, the matrix of proportional gains is $K_\Phi = \mathrm{diag}(k_\phi,k_\theta,k_\beta)$. The commands that are unachievable due to actuator constraints are avoided by passing the reference input through the first-order lag pre-filters,
\begin{equation}\label{eq:68}
  \dot{x}_{pf} = \frac{1}{\tau_{pf}}\left(x_{ref} - x_{pf}\right)
\end{equation}
\noindent
where $\tau_{pf}\in\{\tau_\phi,\tau_\theta,\tau_\beta\}$. The pre-filter time-constants are chosen by a trade-off between tracking performance and command saturation. Tuned gains of angle-loop and pre-filter time-constants are listed in Table \ref{tab:LoopGains}.

\renewcommand\arraystretch{1.3}
\begin{table}
\centering
\caption{Controller gains and pre-filter time-constants for XQ-6B}
\label{tab:LoopGains}
\arrayrulecolor{black}
\begin{tabular}{ccc}
\toprule
\multicolumn{2}{c}{Controller Gains} & \multirow{2}{*}{\begin{tabular}[c]{@{}c@{}}Pre-Filter\\Time Constants\end{tabular}}  \\  \cmidrule(rl){1-2}
Rate-Loop & Angle-Loop   & \\ \cmidrule(rl){1-3}
$k_p = 10.0$ & $k_\phi = 2.0$   & $\tau_\phi = 0.7$ \\
$k_q = 10.0$ & $k_\theta = 2.0$ & $\tau_\theta = 1.0$ \\
$k_r = 10.0$ & $k_\beta = 2.0$  & $\tau_\beta = 0.7$ \\
\bottomrule
\end{tabular}
\arrayrulecolor{black}
\end{table}
\renewcommand\arraystretch{1.0} 

\section{Simulation Results}
In this section, the performance of proposed PMLR based nonlinear control allocation is evaluated. Simulation of the complete closed-loop control system is developed in the MATLAB/Simulink$^\circledR$. Aircraft dynamics are based on the 6-DOF nonlinear model of the aircraft. The performance of nonlinear control allocation is compared for polynomial and PMLR based control surface models using a U-turn maneuver. To generate such a maneuver, a square pulse of 50 deg roll angle command is tracked, whereas, pitch angle is regulated at its initial trim value (3.73 deg) and, sideslip angle is regulated at zero. At the start of maneuver, the aircraft is initially trimmed at a velocity of 25 m/s and an altitude of 500 m. Reason for selection of such maneuver is that a sizable control demands can be produced in all three rotary axes of the aircraft body, which in turn force all control surfaces to sweep a large part of deflection range.

Control allocation performance is evaluated for two different innovative effector configurations:
\begin{enumerate}
  \item \emph{Split-aileron – Ruddervator} Configuration
  \item \emph{Elevator – Rudderon} Configuration
\end{enumerate}

Let us express the local control effectiveness matrix in a partitioned form

\begin{equation}\label{eq:69}
  G = q_\infty S L_{ref} \HVEC{M_1}{|}{M_2}
\end{equation}

where $M_1\in\R{3\times\kappa_1}$ and $M_2\in\R{3\times\kappa_2}$ are two partitions which are defined according to the effector configuration under consideration.

In \emph{Split-aileron – Ruddervator} configuration, all four clamshell control surfaces are ganged together to work as a split-ailerons, whereas, all flaps work as ruddervators. This ganging scheme is expressed as follows:

\begin{equation}\label{eq:74}
\delta_a  = \frac{1}{2}\left[ \left(\delta_{7L}-\delta_{8U}\right) + \left(\delta_{7U}-\delta_{8L}\right) \right]
\end{equation}

\noindent
where $\delta_a$ is the aileron command. Any two diagonal clamshell surfaces are deflected at a time depending upon the polarity of the aileron command, which is defined such that a positive $\delta_a$ produces a positive rolling moment and vice versa. This ganging scheme also reduces total number of control-effectors to seven. Thus, the partitions of control effectiveness matrix become,

\begin{equation}\label{eq:75a}
  M_1 = \begin{bmatrix} \s_{\delta_1} & \s_{\delta_2} & \cdots & \s_{\delta_6} \end{bmatrix}
\end{equation}

\begin{equation}\label{eq:75b}
M_2 = \begin{cases}
     -\s_{\delta_{8L}} + \s_{\delta_{7U}} & \delta_a < 0 \\
      \s_{\delta_{7L}} - \s_{\delta_{8U}} & \delta_a \geq 0
   \end{cases}
\end{equation}

In this configuration, clamshell surfaces are restricted to produce only rolling moment, whereas, no such restriction is applied to the flaps. Since, the clamshells surfaces become main contributors of the commanded rolling moment, the flaps are forced to contribute pitching and yawing moments by the control allocation method. The residual rolling moment of the flaps is also compensated by the clamshell surfaces through control allocation.

In \emph{Elevator – Rudderon} configuration, all six flaps are ganged together to work as a single elevator. Moreover, all four clamshell control surfaces are operated independently to predominantly effect in roll- and yaw- axes. The elevator ganging scheme is expressed as follows:

\begin{equation}\label{eq:76}
\delta_e = \frac{1}{6}\sum_{i=1}^{6}\delta_{i}
\end{equation}

This ganging scheme reduces total number of control-effectors to five. Thus the partitions of control effectiveness matrix become,

\begin{equation}\label{eq:77}
  M_1 = \sum_{i=1}^{6}\s_{\delta_i}
\end{equation}

\begin{equation}\label{eq:78}
M_2 = \begin{bmatrix}
\s_{\delta_{7L}} & \s_{\delta_{7U}} & \s_{\delta_{8L}} & \s_{\delta_{8U}} \end{bmatrix}
\end{equation}

In this configuration, the flaps are restricted to produce only pitching moment, whereas, no such restriction is applied to the clamshell surfaces. Since, the flaps become main contributors of the commanded pitching moment, the clamshell surfaces are forced to contribute rolling and pitching moments by the control allocation method. The residual pitching moment of the clamshell surfaces are also compensated by the flaps through control allocation.

\begin{figure}
  \centering
  \includegraphics[width=0.95\linewidth]{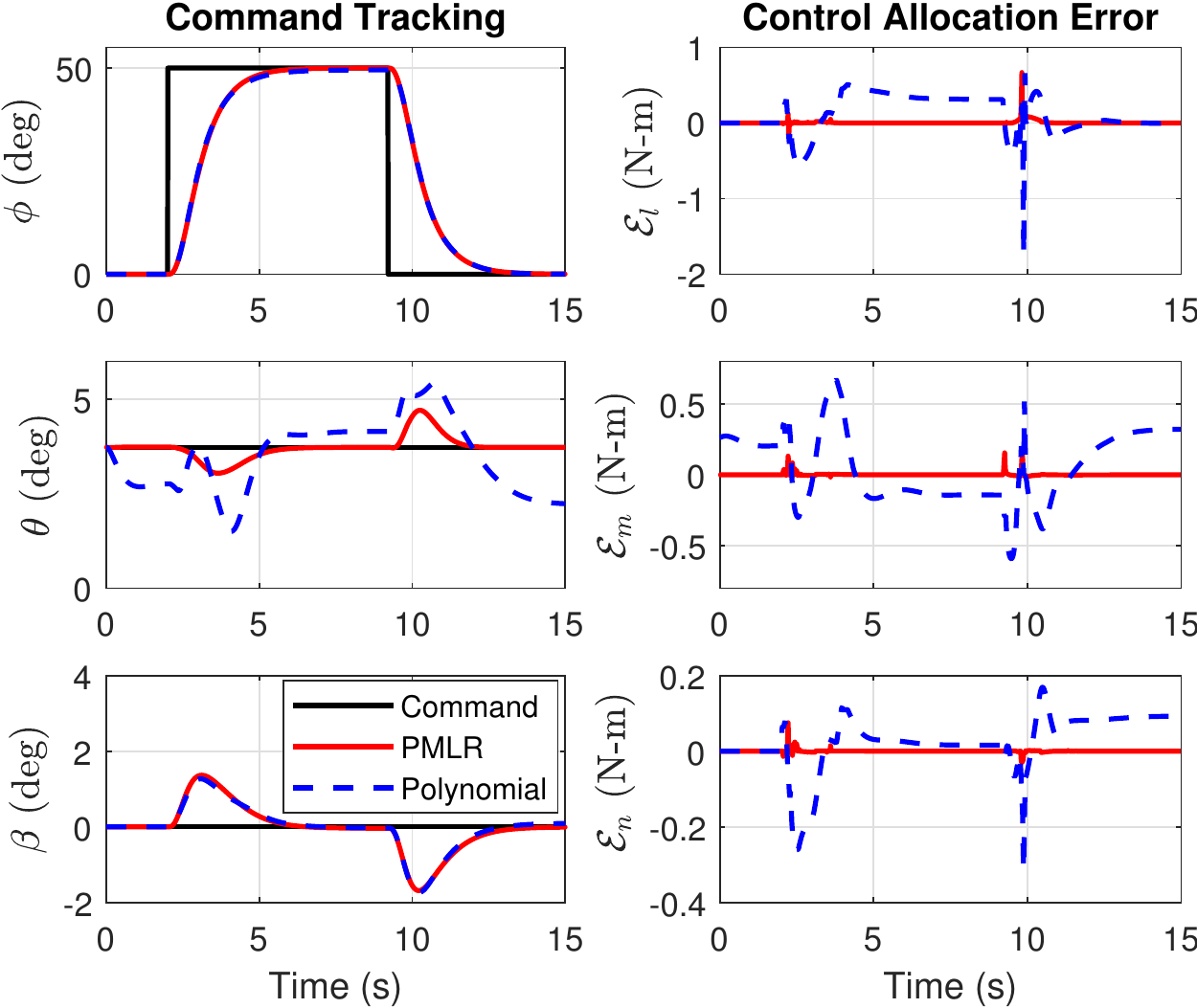}
  \caption{Simulation Results for \emph{Split-aileron – Ruddervator} Configuration} \label{fig:ResultsConf2}
\end{figure}

The performance is evaluated in terms of the control allocation error, $\mathcal{E}(t)$, which is a difference between the vector of demanded control moments, $T_{dem}(t)$, and the vector of actual control moments delivered by the control surface deflections, $T_\delta(t)$, which is expressed as

\begin{equation}\label{eq:70}
  \mathcal{E}(t) \equiv \begin{bmatrix} \mathcal{E}_l \\ \mathcal{E}_m \\ \mathcal{E}_n \end{bmatrix} = T_{dem}(t) - T_\delta(t)
\end{equation}

\begin{figure}
  \centering
  \includegraphics[width=0.95\linewidth]{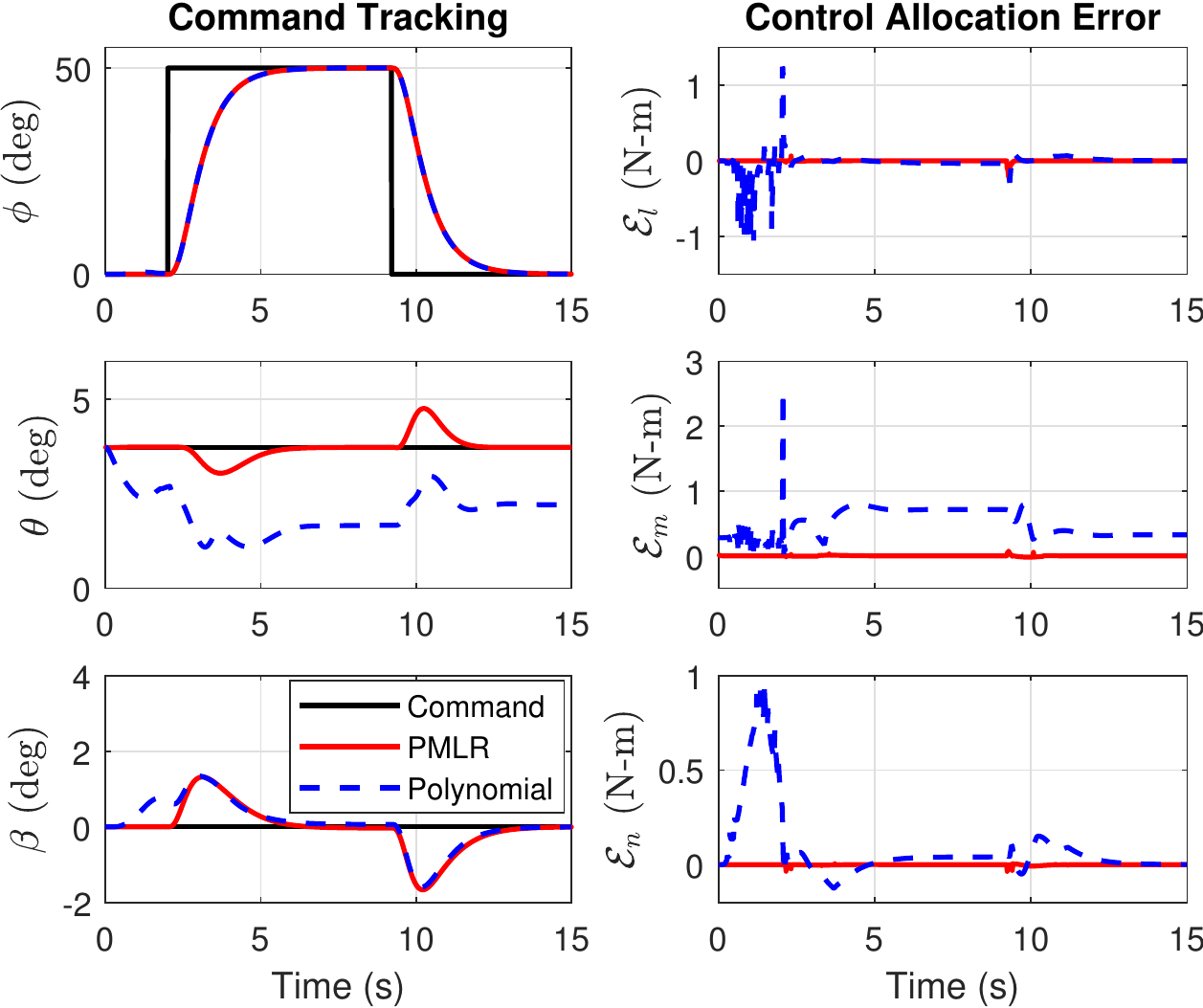}
  \caption{Simulation Results for \emph{Elevator – Rudderon} Configuration} \label{fig:ResultsConf3}
\end{figure}

Simulation results for both configurations are shown in Fig. \ref{fig:ResultsConf2} and Fig. \ref{fig:ResultsConf3}. The comparison of control allocation performance, summarized in terms of RMS of control allocation errors, is given in Table \ref{tab:ErrorsConf}. In general, due to greater modeling accuracy, the  PMLR-based control allocation produces much lower errors as compared to polynomial-based control  allocation. For the both configurations, 
the polynomial-based control allocation results in larger inaccuracies in all three control channels, thus, control performance degradation is observed. On the other hand, the  PMLR-based control allocation gives equally accurate results for both configurations, which results in superior control performance.

\renewcommand\arraystretch{1.2}
\begin{table}
\centering
\caption{Control Allocation Errors}
\label{tab:ErrorsConf}
\begin{tabular}{ccccc}
\toprule
& &  & RMS $\mathcal{E}$ [N-m] &  \\ \cmidrule(rl){3-5}
\multirow{-2}{*}{Conf.} & \multirow{-2}{*}{Method} & Roll ($\mathcal{E}_l$) & Pitch  ($\mathcal{E}_m$) & Yaw  ($\mathcal{E}_n$) \\ \cmidrule(rl){1-5}
& PMLR & $0.03$ &  $0.010$ &  $0.004$ \\
\multirow{-2}{*}{1} & Polynomial & $0.27$ &  $0.250$ &  $0.080$ \\ \cmidrule(rl){2-5}
& PMLR & $0.01$ &  $0.007$ &  $0.003$ \\
\multirow{-2}{*}{2} & Polynomial & $0.14$ &  $0.530$ &  $0.210$ \\
\bottomrule
\end{tabular}
\end{table}
\renewcommand\arraystretch{1.0}

\section{Conclusion}
A nonlinear control allocation method was presented, which uses control surface model based on newly developed piecewise multi-linear representation to compute control commands. Aerodynamic data of an aircraft’s control surfaces is usually a piecewise multi-linear function of states and control surface deflections. Thus, it can be modelled exactly using the proposed piecewise multi-linear representation. It was shown that the special structure of PMLR makes it possible to compute the locally affine approximation of nonlinear moment versus deflection relationships through explicit formulation. Consequently, a PMLR-based model directly fits into computationally efficient affine control allocation framework for solving the nonlinear control allocation problem.

\section*{Acknowledgment}
Authors are thankful to Professor Dr. Zang Weiguo, Dr. Qu Xiaobo and Dr. Shi Jingping (School of Automation, Northwestern Polytechnical University, Xi’an) for their technical and moral support.

\ifCLASSOPTIONcaptionsoff
  \newpage
\fi



%
\bibliographystyle{IEEEtran}
\bibliography{References}

\begin{IEEEbiography}[{\includegraphics[width=1in,height=1.25in,clip,keepaspectratio]{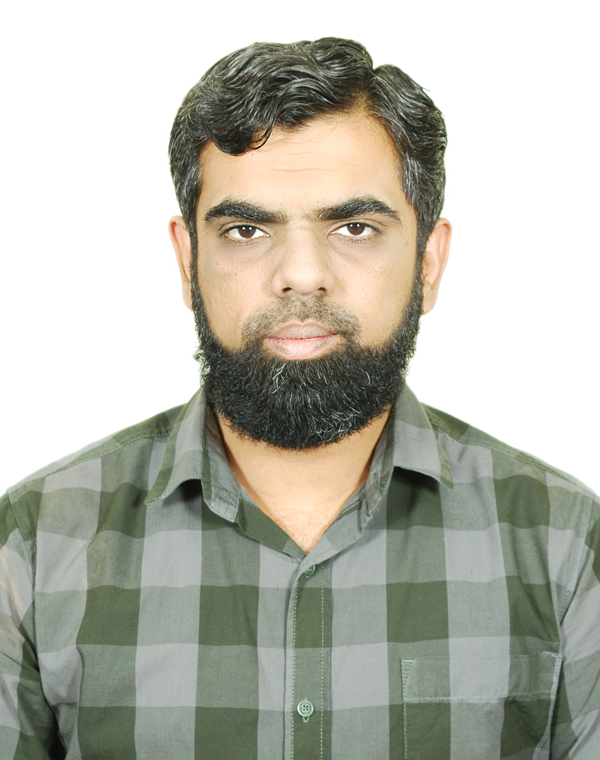}}]{Jahanzeb Rajput}
received the B.E. degree in Electronics Engineering from Mehran University of Engineering and Technology, Jamshoro, Pakistan, in 2005. He received the M.S. degree in Control Engineering from University of Engineering and Technology, Lahore, Pakistan, in 2007. He received the Ph.D. degree in Navigation, Guidance and Control from Northwestern Polytechnical University, Xi’an, China, in 2016. His main research interests include nonlinear control, control allocation, fault-tolerant control, reconfigurable control, modeling, identification and simulation.
\end{IEEEbiography}



\begin{IEEEbiography}[{\includegraphics[width=1in,height=1.25in,clip,keepaspectratio]{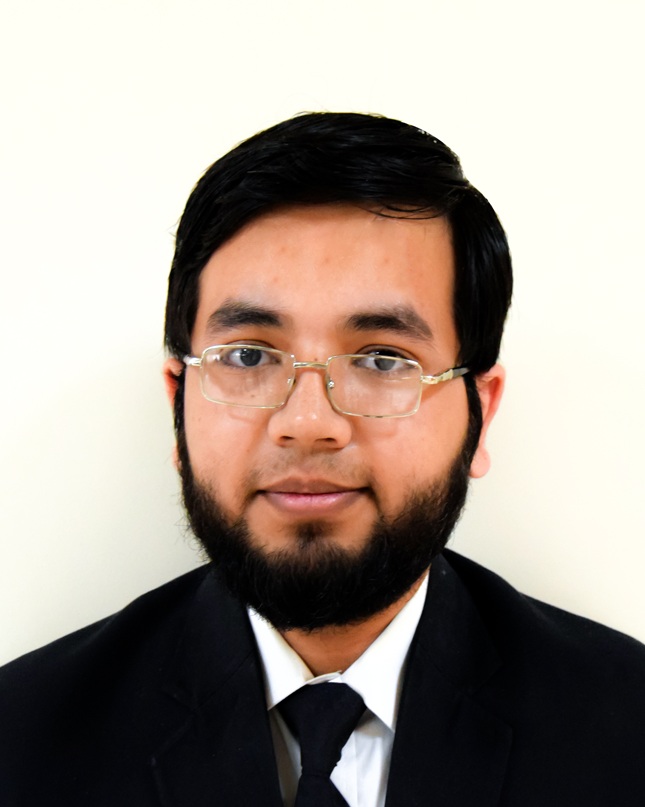}}]{Hafiz Zeeshan Iqbal Khan}
received his BS degree magna cum laude in Aerospace Engineering, from Institute of Space Technology, Islamabad, Pakistan, in 2015, with honors and President of Pakistan gold medal. He received his MS degree summa cum laude in Aerospace Engineering specializing in Guidance, Navigation \& Control, from Institute of Space Technology, Islamabad, Pakistan, in 2019, with honors. His main research interests include dynamics \& control of aerospace vehicles, robust control, nonlinear control, geometric control, and machine learning.
\end{IEEEbiography}
\vfill



\end{document}